%% file: main.tex
\documentclass{article}

\PassOptionsToPackage{numbers, compress}{natbib}
 \usepackage[preprint]{paper_style}

\usepackage[utf8]{inputenc} % allow utf-8 input
\usepackage[T1]{fontenc}    % use 8-bit T1 fonts
\usepackage{hyperref}       % hyperlinks
\usepackage{url}            % simple URL typesetting
\usepackage{booktabs}       % professional-quality tables
\usepackage{amsfonts}       % blackboard math symbols
\usepackage{nicefrac}       % compact symbols for 1/2, etc.
\usepackage{microtype}      % microtypography
\usepackage{graphicx}
\usepackage{arydshln}
\usepackage{multirow}
\usepackage{subfig}
\usepackage{bm}
\usepackage{amsmath}

\usepackage[usenames,dvipsnames]{xcolor}

\newcommand{\attn}{\textcolor{red}}
\usepackage{csquotes}

\newcommand{\beginsupplement}{%
        \setcounter{table}{0}
        \renewcommand{\thetable}{S\arabic{table}}%
        \setcounter{figure}{0}
        \renewcommand{\thefigure}{S\arabic{figure}}%
     }

\title{
Countering hate on social media: Large scale classification of hate and counter speech}

\author{%
Joshua Garland\thanks{Denotes equal contribution.}\\
Santa Fe Institute\\
Santa Fe, NM 87501 USA\\
\texttt{joshua@santafe.edu} \\
\And
Keyan~Ghazi-Zahedi$^*$ \\
Max Planck Institute for \\ Mathematics in the Sciences\\
Inselstrasse 22, 04103\\
Leipzig, Germany \\
\AND
Jean-Gabriel Young\\
Center for the Study of Complex Systems\\
University of Michigan\\
Ann Arbor, MI 48109, USA
\AND
Laurent~Hébert-Dufresne\\
Vermont Complex Systems Center\\
University of Vermont \\Burlington, VT 05405, USA
\And
Mirta~Galesic \\
Santa Fe Institute \\
Santa Fe, NM 87501 USA
}

\begin{document}

\maketitle

\begin{abstract}
Hateful rhetoric is plaguing online discourse, fostering extreme societal movements and possibly giving rise to real-world violence. A potential solution to this growing global problem is citizen-generated counter speech where citizens actively engage in hate-filled conversations to attempt to restore civil non-polarized discourse. However, its actual effectiveness in curbing the spread of hatred is unknown and hard to quantify. One major obstacle to researching this question is a lack of large labeled data sets for training automated classifiers to identify counter speech. Here we made use of a unique situation in Germany where self-labeling groups engaged in organized online hate and counter speech. We used an ensemble learning algorithm which pairs a variety of paragraph embeddings with regularized logistic regression functions to classify both hate and counter speech in a corpus of millions of relevant tweets from these two groups. Our pipeline achieved macro F1 scores on out of sample balanced test sets ranging from 0.76 to 0.97---accuracy in line and even exceeding the state of the art. On thousands of tweets, we used crowdsourcing to verify that the judgments made by the classifier are in close alignment with human judgment. We then used the classifier to discover hate and counter speech in more than 135,000 fully-resolved Twitter conversations occurring from 2013 to 2018 and study their frequency and interaction. Altogether, our results highlight the potential of automated methods to evaluate the impact of coordinated counter speech in stabilizing conversations on social media.
\end{abstract}

\section{Introduction}
\label{sec:intro}

Hate speech is a growing problem in many countries \cite{bakalis2015,hawdon2017exposure} as it can have serious psychological consequences \cite{oksanen2018perceived} and is related to, and perhaps even contributing to, real-world violence \cite{muller2019fanning}. While censorship can help curb hate speech \cite{alvarez2018normative}, it can also impinge on civil liberties and might merely disperse rather than reduce hate \cite{chandrasekharan2017you}. A promising approach to reduce toxic discourse without recourse to outright censorship is so-called \emph{counter speech}, which broadly refers to citizens' response to hateful speech in order to stop it, reduce its consequences, and discourage it \cite{benesch2016considerations,rieger2018hate}.

It is unknown, however,  whether counter speech is actually effective  due to the lack of systematic large-scale studies on its impact \cite{gaffney2019cyberbullying,gagliardone2015countering}. A major reason has been the difficulty of designing automated algorithms for discovering counter speech in large online corpora, stemming mostly from the lack of labeled training sets including both hate and counter speech. Past studies that provided insightful analyses of the effectiveness of counter speech mostly used hand-coded examples and were thus limited to small samples of discourse \cite{mathew2018analyzing,mathew2019thou,wright2017vectors,ziegele2018journalistic}. 

The first step in understanding the effectiveness of counter speech is to classify that speech. 
We perform the first large scale classification study of hate and counter speech by using a unique situation in Germany, where self-labeling hate and counter speech groups engaged in discussions around current societal topics such as immigration and elections. One is ``Reconquista Germanica" (RG), a highly-organized hate group which aimed to disrupt political discussions and promote the right-wing populist, nationalist party Alternative f\"ur Deutschland (AfD). At their peak time, RG had between 1,500 and 3,000 active members.
The counter group  ``Reconquista Internet'' (RI)
formed in late April 2018 
with the aim of countering RG's hateful messaging through counter speech and to re-balance the public discourse. Within the first week, approximately 45,000 users joined the discord server where RI was being organized. At their peak, RI had an estimated 62,000 registered and verified members, of which over 4,000 were active on their discord server for the first few months. However, RI has quickly lost a significant amount of active members, splintering into independent though cooperating smaller groups.  We collected millions of tweets from members of these two groups and built labeled training set orders of magnitude larger than existing data sets \citet{mathew2018analyzing,wright2017vectors,ziegele2018journalistic,ziems2020racism}. By building an ensemble learning system with this large corpus we were able to train highly accurate classifiers which matched human judgment. We were also able to use this system to study more than 130,000 conversations between these groups to begin studying the interactions between counter and hate groups on Twitter---an important first step in studying the impacts of counter speech on a large scale.

\section{Background and Past Research}
\subsection{Hate and counter speech}
There are many definitions of online hate speech and its meaning is developing over time. According to more narrow definitions, it refers to insults, discrimination, or intimidation of individuals or groups on the Internet, on the grounds of their supposed race, ethnic origin, gender, religion, or political beliefs  \cite{blaya2019cyberhate,weber2009manual}. However, the term online hate speech can also be extended to speech that aims to spread fearful, negative, and harmful stereotypes, call for exclusion or segregation, incite hatred, and encourage violence against a particular group \cite{gagliardone2015countering,youtube_guide,twitter_guide,facebook_guide}, be it using words, symbols, images, or other media.

Counter speech entails a citizen generated response to online hate in order to stop and prevent the spread of hate speech, and if possible change perpetrators' attitudes about their victims. Counter speech intervention programs focus on empowering Internet users to speak up against online hate \cite{gagliardone2015countering}. For instance, programs such as \textit{seriously} \cite{seriously-website}
and the \textit{Social Media Helpline} \cite{smh-website} 
help users to recognize different kinds of online hate and prepare appropriate responses. Counter speech is seen as a feasible way of countering online hate, with a potential to increase civility and deliberation quality of online discussions \cite{ziegele2018journalistic,habermas2015between}. 

\subsection{Classification of Hate and Counter Speech}
There has been a lot of work on developing classifiers to detect hate 
speech online (e.g., \cite{brassard2018impact,burnap2015detecting,
burnap2016us,ribeiro2018characterizing,zhang2019hate,bosco2018overview,
gibert2018hate,kshirsagar2018predictive,macavaney2019hate,malmasi2018challenges,pitsilis2018effective,al2019detection,
vidgen2020detecting,zimmerman2018improving}).  Many different learning 
algorithms have been used to perform this classification, ranging from support vector machines and random forests to convolutional and 
recurring neural networks \cite{zhang2019hate,burnap2016us,
bosco2018overview,gibert2018hate,kshirsagar2018predictive,
malmasi2018challenges,pitsilis2018effective,al2019detection,
vidgen2020detecting,zimmerman2018improving}). 
These algorithms use a variety of feature extraction methods, for example, frequency scores of different n-grams, 
 word and document embeddings \cite{doc2vec,pennington2014glove}, sentiment scores \cite{brassard2018impact,burnap2015detecting}, part-of-speech scores such as the frequency of adjectives versus nouns used to describe target groups 
 `othering' language (e.g., 'we' vs. 'them'  \cite{burnap2016us}), and meta-information about the text authors (e.g., keywords from user bios, usage patterns, their connections based on replies, retweets, and following patterns  \cite{ribeiro2018characterizing}). \citet{zhang2019hate} compare several state-of-the art methods for automatic detection of hate speech, including SVM and different implementations of convolutional neural networks (CNN) using word embeddings, on seven different Twitter data sets. The best performing methodology, based on a combination of CNN and gated recurrent networks (GRU), yields macro F1 scores ranging from 0.64 to 0.83. Other promising approaches include an ensemble of different CNNs with different weight initializations proposed by Zimmerman et al. \cite{zimmerman2018improving}, BERT \cite{devlin2018bert}, and a multi-view stacked SVM approach proposed by McAvaney et al. \cite{macavaney2019hate}. The best results reported for these approaches are achieved by \citet{macavaney2019hate} using the \texttt{hatebase.org} database (a set of 24,802 tweets provided by \citet{davidson2017automated}, receiving F1 scores of 0.91 with a neural ensemble and 0.89 using BERT.

Compared to the number of studies investigating automatic detection of online hate, there have been far fewer studies that aim to automatically detect counter speech. One reason for this is the difficulty and subjectivity 
of automated identification of counter speech \cite{kennedy2017technology}.  As a result, most past studies use hand-coded examples for this task. For instance, \citet{mathew2019thou} analyzed more than 9,000 hand-coded counter speech and neutral comments posted in response to hateful YouTube videos. They found that for discriminating counter speech vs. non-counter speech, the combination of tf-idf vectors as features and logistic regression as the classifier performs best, achieving F1 score of $0.73$. In another study, \citet{mathew2018analyzing} analyzed 1,290 pairs of Twitter messages containing hand-coded hate and counter speech. In this data set, a boosting algorithm based mostly on tf-idf values and lexical properties of tweets performed best, achieving F1 score of $0.77$. \citet{wright2017vectors} provide a qualitative analysis of individual examples of counter speech. \citet{ziegele2018journalistic} employed 52 undergraduate students to hand-code 9,763 Facebook messages. A study concurrent to ours \cite{ziems2020racism} investigated hate and counter speech in the context of racist sentiment surrounding COVID-19. They hand-coded 2,319 tweets, of which they labeled 678 as hateful, 359 as counter speech, and 961 as neutral. They were able to achieve F1 scores on unbalanced sets of 0.49 for counter and 0.68 for hate. 

While extremely useful as a first step in analyzing counter speech, these studies have limited applicability because manual coding of counter speech is costly, hard to scale to the size needed to train sophisticated classifiers, and a task of considerable difficulty for the manual coders.

\input{methods.tex}

\input{results.tex}

\section{Conclusion}
\label{sec:concl}

Online hate speech is a problem shared by every social media platform, and yet there are still no clear solutions to this growing problem. A potential solution aimed at returning online discourse to civility is citizen generated counter speech. Until now, studying counter speech and its effectiveness has been limited to small-scale hand-labeled studies. In this paper, we leveraged a unique situation in Germany to perform the first large scale automated classification of counter speech. Our methods provided F1 scores on a balanced set of 500,000 out-of-sample tweets ranging from 0.76 to 0.97 depending on the confidence threshold being used. Beyond accuracy measures, we used crowdsourcing to verify that the conclusions reached by our classifier were in-line with human judgment. 

We were able to use this classification algorithm to identify hate and counter speech in over 135,000 fully resolved Twitter conversations from 2013-2018. 
Our results suggest that counter speech contributed to depolarization of discussions and that organized counter speech by RI might have stimulated further counter speech and attracted less hateful responses. Organized counter speech may therefore be a powerful solution to combating the spread of hate online. We hope that the framework developed in this paper will be a starting point to understand the dynamics between hate and counter speech and help develop actionable strategies.

\section*{Acknowledgments}
 The authors would like to thank Will Tracy and Santa Fe Institute's Applied Complexity team for support and resources throughout this project. J.G. was partially supported by an Omidyar and an Applied Complexity Fellowship at the Santa Fe Institute. J.-G.Y. was supported by a James S. McDonnell Foundation Postdoctoral Fellowship Award. L.H.-D. was supported by Google Open Source under the Open-Source Complex Ecosystems And Networks (OCEAN) project. Any opinions, findings, and conclusions or recommendations expressed in this material are those of the authors and do not necessarily reflect the views of Google Open Source.
 M.G. was partially supported by NSF-DRMS 1757211. 
% 
% \bibliography{references}

\input{ref.bbl}
\input{supp.tex}

\end{document}

%% file: methods.tex
\section{Data and Methods}
\label{sec:methods}

\subsection{Data Collection Strategy}
\label{sec:data}
To train our classification algorithms, we collected more than 9 million relevant tweets. Of these tweets, we labeled 4,689,294 as originating from a hate account (RG member tweets) and 4,323,881 as originating from a counter speech account (RI member tweets). 

We built the initial corpus of hate speech by downloading the complete timelines of 2,120 known members of RG using the Twitter API. We further verified these accounts by ensuring that the names and/or bios of these accounts contained known RG badges (see Supplementary Table~\ref{tab:summary} for a list of these features) and no known badges of RI. This resulted in more than 4.6 million tweets which most likely contained hateful rhetoric. Of course, not all tweets by these accounts were necessarily hateful messages,  but active members of RG used their public accounts to either spread hateful rhetoric, promote alt-right propaganda, or engage in directly hateful speech.
Therefore, we considered the tweets sent from these accounts to be largely representative of hateful speech.

Whereas hate accounts were quite abundant and often willing to self identify, RI members proved more challenging to label. We began our search with a hand-curated list of 103 known RI members and scraped the most recent timelines of these limited but known members of RI. This did not give us a large or diverse enough corpus of counter speech; especially since many of these accounts do not solely focus on counter speech. 
Therefore we also scraped the follower-followee network of these 103 RI members using the Twitter API. 
 
 We then constructed a list of potential counter accounts in these networks, including those that appeared in at least 5 of the follower-followee networks of known RI members. This resulted in a list of 70,537 potential accounts engaging in counter speech. From these accounts we retained only those that used language features typical of RI members in their bios (see Table~\ref{tab:summary} for a list of these features). 
 We then further restricted these potential accounts by eliminating any users from this subset which contained any RG features. This was an attempt to eliminate troll accounts which had both hate and counter features present in their bios. This process resulted in a total of 1,472 accounts which we labeled as counter accounts. 
 
 To build our corpus of counter speech we collected the timelines of each of these accounts if they were publicly available. This resulted in a total of 4,323,881 tweets which had a high probability of containing counter speech. Similarly to hate accounts, and perhaps more so in the case of counter accounts, not all tweets they produced can be considered as counter speech. We handled this challenge directly in the classification pipeline, discussed in Section~\ref{sec:poe}.

\begin{figure}%
    \centering
    \subfloat{{\includegraphics[width=.3\textwidth]{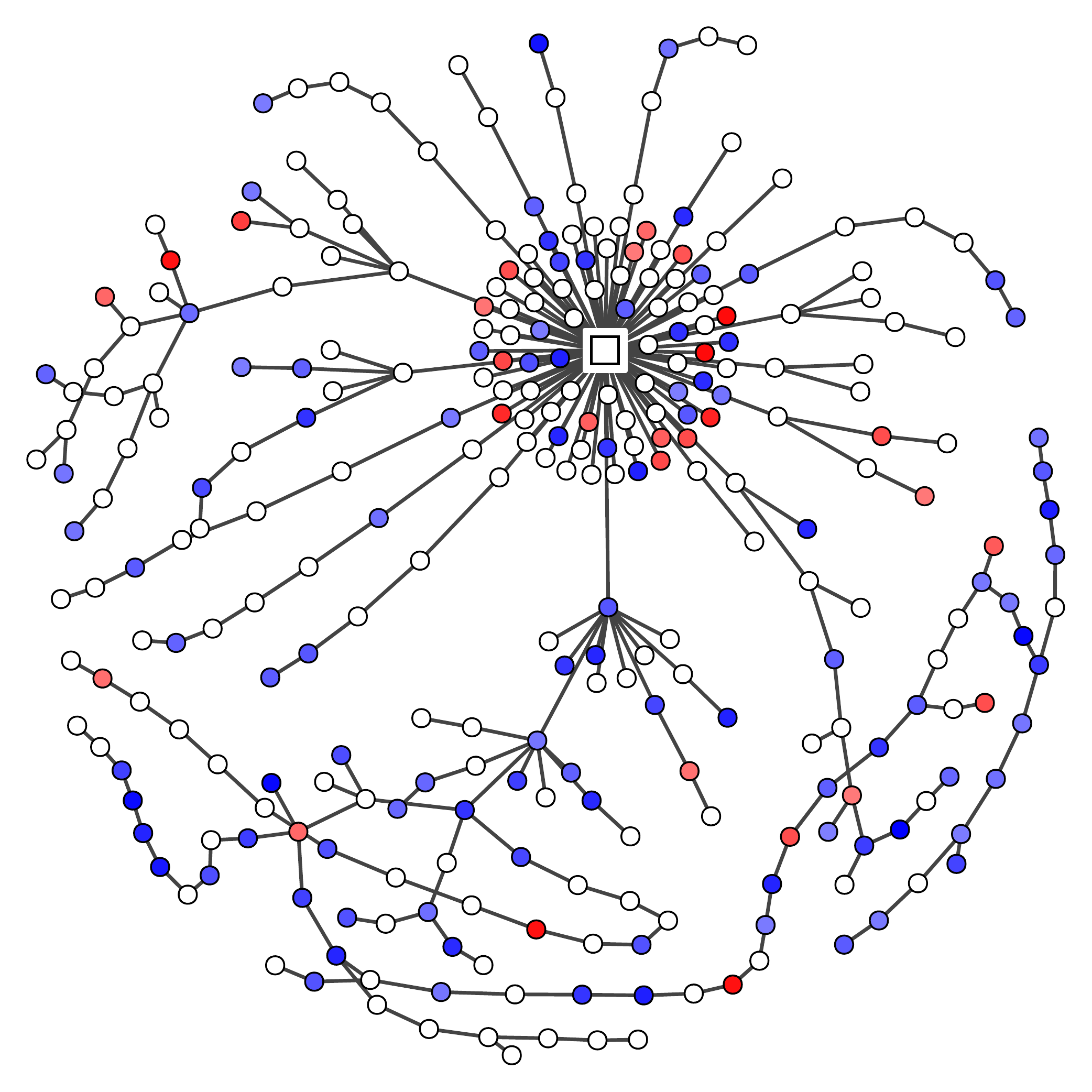} }}%%
    \subfloat{{\includegraphics[width=.3\textwidth]{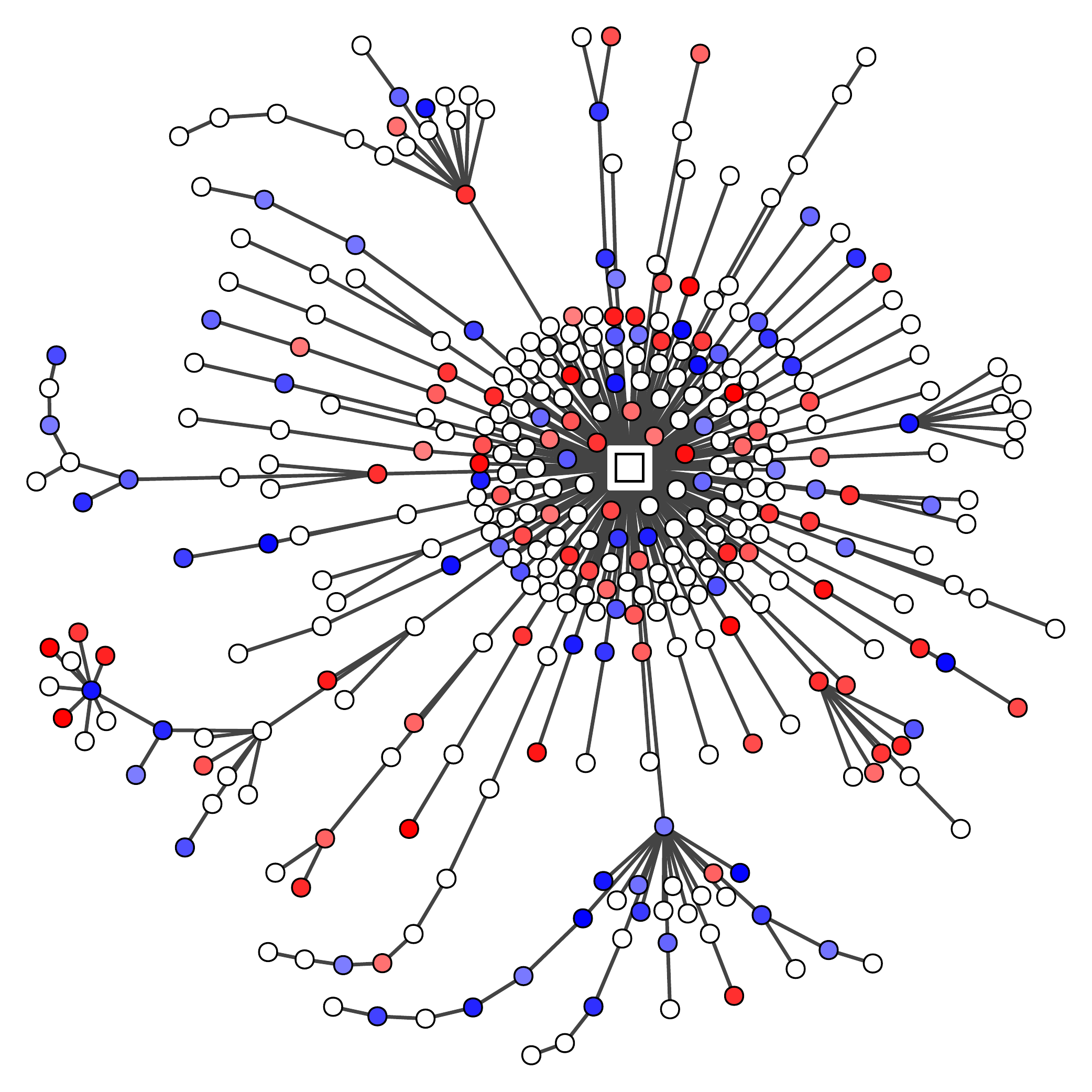} }}%
    \subfloat{{\includegraphics[width=.3\textwidth]{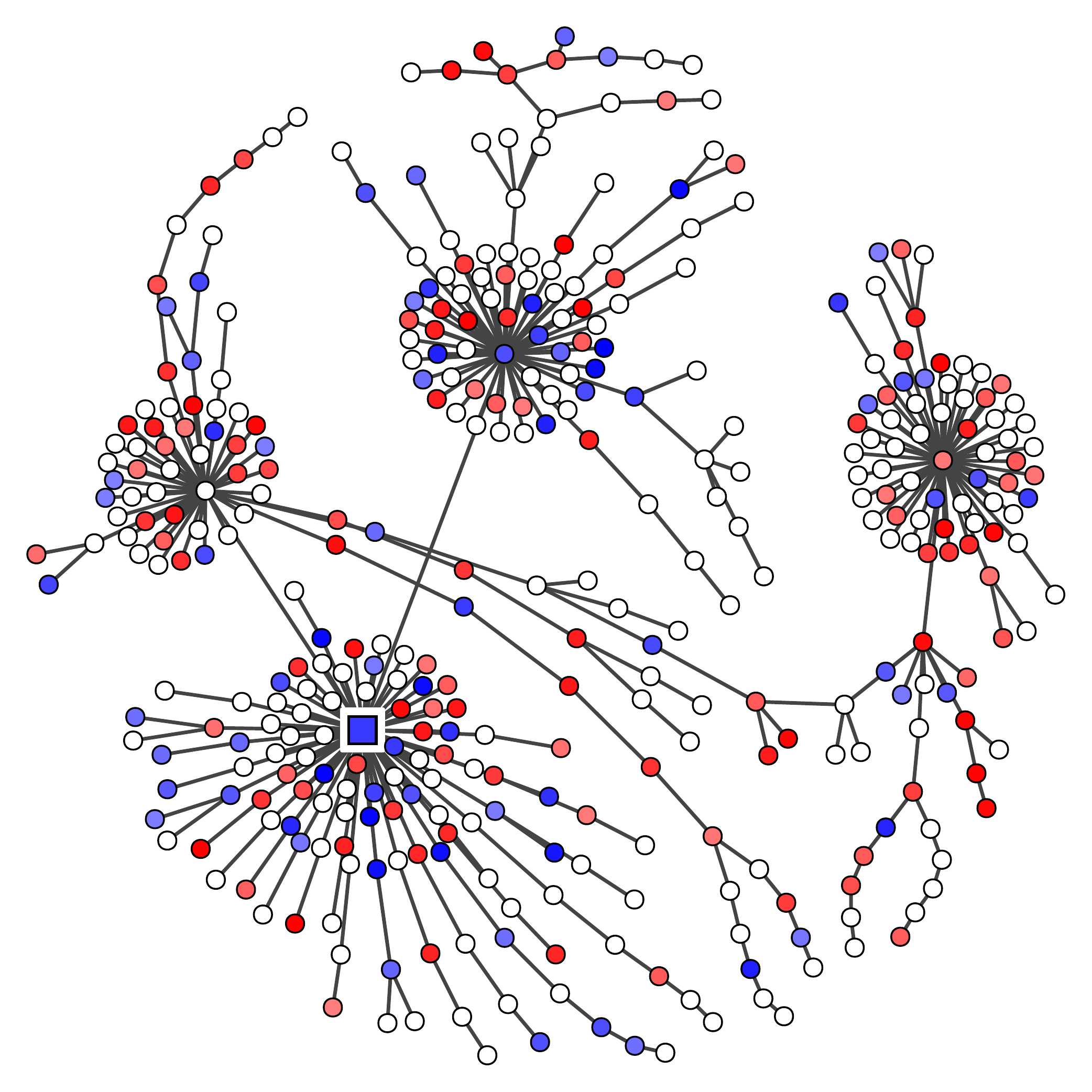} }}
    \caption{Examples of Twitter conversations (reply trees) with labeled hate (red), counter (blue), and neutral speech (white). The root node is shown as a large square. We used a confidence threshold of $\gamma=0.75$ and a panel of 25 experts to classify these tweets, as described later in Section~\ref{sec:poe}.  }%
    \label{fig:coloredtrees}%
 \end{figure}

 Additionally, we collected 204,544 conversations (reply trees) that grew in response to tweets of prominent accounts engaged in political speech on German Twitter from 2013 to 2018. These included accounts of large news organizations (e.g., faznet, tagesschau, tagesthemen, derspiegel and spiegelonline, diezeit, and zdfheute), well-known journalists and bloggers (e.g., annewilltalk, dunjahayali, janboehm, jkasek, maischberger, nicolediekmann), and politicians (e.g. cem\_oezdemir, c\_lindner, goeringeckardt, heikomaas, olafscholz, renatekuenast), all of which were known to be targets of hate speech. Indeed, the majority of these conversations involve instances of both hate and counter speech.  To compare trends in hate and counter speech over time, we focused on 137,725 trees which originated from 11 accounts that contributed trees in at least 69 of 72 possible months throughout the examined period: derspiegel, goeringeckardt, jkasek, olafscholz, regsprecher, zdfheute, c\_lindner, faznet, janboehm, nicolediekmann, and tagesschau. While the 1,062,267 tweets contained in these trees were not used during the classification training pipeline, some of these tweets appeared in RG and RI timelines. As such, care was taken as to not have any classifiers evaluate tweets that were previously seen during training. Figure~\ref{fig:coloredtrees} shows a few example trees labeled using the pipeline described in Section~\ref{sec:classification}.

\subsection{Classification Pipeline}
\label{sec:classification}
As is common in the literature  \cite{schmidt2017survey} we split our classification pipeline into two stages: extraction of features from text, and classification based on those features. Before tweets were used in this pipeline they went through a minor preprocessing stage. All of the text was made lower case, hashtags, usernames e.g., @username, punctuation and ``RT:'' were all stripped out of the tweet's text. Finally, depending on the model being trained, we removed stop words depending on one of two lists (``heavy" and ``light"), or we did not remove any stop words. The ``heavy" stop word list eliminated 231 German words based on nltk's German stop word list. The ``light" stop word list was based on the heavy list without all words which have been shown to be relevant identifiers in an ``us vs. them'' discourse \cite{burnap2016us}, e.g., wir, uns, sie (we, us, them). This list eliminated 48 words (see Supplementary Material).

To extract features from each processed timeline tweet, we constructed paragraph embeddings, also known as doc2vec models \cite{doc2vec}, using the standard gensim implementation \cite{rehurek_lrec}. We will refer to a generic doc2vec model as $\mathcal{M}_{d2v}$. We performed a parameter sweep following standard practice and the guidelines of \cite{lau2016empirical}. This sweep includes the analysis of several doc2vec parameters e.g, maximum distance between current and predicted words, ``distributed-memory'' vs ``distributed bag of words'' frameworks, and five different document label types, as well as three levels of stop word removal.

The five different document label types were as follows. 1) Each tweet was treated as a single document and labeled with a \emph{unique} label. 2) All tweets by a single \emph{author} used the same document label. This effectively made every tweet by a particular user a single document. These were the more traditional choices for document labeling. We also used three other labels which incorporate the classification stage into the feature development: 3) Each tweet from RG was labeled ``hate" and each tweet from RI was labeled ``counter", we call this the \emph{group} label. This treats all RG tweets as one document and all RI tweets as another document. While this incorporates the label we care about into the feature development stage it conflates all the tweets into two documents. To avoid this we also trained $\mathcal{M}_{d2v}$ using multiple label setups. In particular, we trained models where we 4) labeled each tweet using both the author's identifier as well as the group identifier and separate models which 5) labeled each tweet with a unique identifier as well as the group identifier.

Every $\mathcal{M}_{d2v}$ was trained on five different but partially overlapping training sets (approximately 27\% overlap). Each training set included 500,000 randomly selected tweets originating from RG accounts and another 500,000 coming from RI accounts. This produced a balanced training set with 50\% hate speech and 50\% counter speech. This is important in interpreting our classification results correctly, and avoiding accuracy inflation due to unbalanced sets, an apparent frequent problem with much of the current literature where hate speech is highly under sampled \cite{zhang2019hate,macavaney2019hate}. We refer to these training sets as $\mathcal{T}_{in,i}$, to denote the $i^{th}$ in-sample training set.

Let $\{\mathcal{M}_{d2v},\mathcal{T}_{in,i}\}$ be a trained doc2vec model and the corresponding training set it was trained on. For each tweet $t_j\in\mathcal{T}_{in,i}$ we use $\mathcal{M}_{d2v}$ to infer a corresponding feature vector $x_j\in \mathbb{R}^{300}$, as $x_j=\mathcal{M}_{d2v}(t_j)$.
With each tweet mapped to a feature vector we constructed a decision boundary between tweets from RG members and tweets from RI members using regularized logistic regression.
In other words, we wrote the likelihood that tweet $j$ is labeled as coming from an RG/RI account as:
\begin{equation}
    h_\theta(x_j)=g(\theta^Tx_j)\text{, }\qquad g(z)=\frac{1}{1+e^{-z}}.
\end{equation} 
where $\theta\in\mathbb{R}^{300}$ is the vector of feature weights.
Given a set of labels $\mathcal{L}=\{H,C\}$ for all tweets, we then learned the vector $\theta$ that best separate the data by minimizing the loss $-\sum_j \log h_\theta(x_j)$ under an $\ell_2$ regularization constraint $\frac{1}{\lambda} \lvert\lvert \theta \rvert\rvert_2$, where $\lambda$ is a fixed regularization parameter.
We finally solved for $\theta$ using the the LBFGS algorithm as implemented in scikit-learn \cite{scikit-learn}. 

To evaluate the accuracy of the resulting hypothesis function $h_\theta$ we evaluated its predictive accuracy on an out of sample test set denoted $\mathcal{T}_{out,i}$. Each out of sample test set $\mathcal{T}_{out,i}$ consisted of 50,000 tweets from both groups, chosen at random while ensuring that $\mathcal{T}_{out,i}\cap\mathcal{T}_{in,i}=\emptyset$. For each, $\mathcal{M}_{d2v}$, $h_\theta$, $\mathcal{T}_{out,i}$ combination we determined the probability of each class label $l\in\mathcal{L}$ for each $t\in\mathcal{T}_{out,i}$. In particular, for each tweet $t\in\mathcal{T}_{out,i}$ and each label $l\in\{H,C\}$ we computed $h_\theta(\mathcal{M}_{d2v}(t))=p(l|\mathcal{M}_{d2v}(t);\theta)$, where
$p(l|\mathcal{M}_{d2v}(t);\theta)$ denotes the probability that a tweet $t$ has label $l$ when classified with  the feature vector calculated with model $\mathcal{M}_{d2v}$. The accuracy of this prediction was then assessed against the known labels. 

{In addition to logistic regressions, we also used word bias and n-gram based classifiers like those used in  \cite{jaki2019right}, as well as xgboost \cite{chen2016xgboost} with a variety of parameters. However, in both cases the accuracy was worse (only slightly so for xgboost) than the logistic regression experiments reported in Section~\ref{sec:results}, so we omit these details for brevity.}

\subsubsection{An Ensemble Learning Based Classifier}
\label{sec:poe}
Instead of looking for the single optimal  $(\mathcal{M}_{d2v}, h_\theta)$ parameterization (which may not exist) we used an ensemble learning approach to classification by constructing a ``panel of experts." The panel is comprised of $N$ experts  which are defined to be the combination of a feature extraction method $\mathcal{M}_{d2v}$ as well as a classification or hypothesis function $h_\theta$. 
An ensemble learning approach combines multiple hypotheses functions to form a more robust hypothesis jointly which can lead to greater generalizability and increased out-of-sample accuracy.

In this ensemble classification method, each expert is given a tweet in a balanced out-of-sample test set $\mathcal{T}_{out,i}$ and asked to assign to it a probability that it belongs to each class $l\in\mathcal{L}$.
For each tweet $t\in\mathcal{T}_{out,i}$ we computed a hate and counter score, $S_h$ and $S_c$ respectively, in the following way:

\begin{equation}
    S_h(t)=\frac{1}{N}\sum_{i=1}^{N}\mathcal{E}_i(t;H),\qquad S_c(t)=\frac{1}{N}\sum_{i=1}^{N}\mathcal{E}_i(t;C) = 1 - S_h(t), 
\end{equation}
where $\mathcal{E}_i(t;l)$ is the probability that expert $i$ assigns to label $l\in\{H,C\}$ for tweet $t$.

For final classification we then defined a ``confidence threshold'' $\gamma\in[1/2,1]$, and used a confidence voting system with thresholding to assign labels to tweets. If $S_h(t)>\gamma$ then $t$ is labeled $H$, and if $S_c(t)>\gamma$ then $t$ is labeled $C$. If $S_c(t)$ and $S_h(t)$ are both less than the given threshold the tweet is marked as neutral speech and the panel effectively abstains from voting.  This results in some tweets which the panel of experts is not confident in either label being unclassified as hate or counter speech. As not every tweet by members of RI and RG are  counter or hate speech this allowed the panel not to be forced to label a tweet in one of these two categories. 
Additionally, note that $\mathcal{T}_{out,i}\cap\mathcal{T}_{in,j}$ is not empty when $i\not=j$.
As such, if a tweet appeared in the training set of an expert, we withheld its vote to avoid leaking training data.

\subsection{Crowdsourcing}
To test whether the automated classifier corresponds to human judgment, we conducted a crowdsourcing study in which 55 human judges evaluated 5000 randomly selected tweets evenly spread across the whole range of scores $S_h(t) \in [0,1]$. Since our corpus mostly contains German tweets, judges were recruited among members of Mechanical Turk who lived in Germany, Austria, or Switzerland. To qualify, they had to complete a relatively difficult German test item taken from a Goethe Institut's test for B1 German level, which asked them to interpret comments of three individuals about violence in video games. Each tweet was evaluated by 3 different judges, and ranked on a scale of 1 to 5, from ``very likely counter speech'' to ``very likely hate speech,`` with 3 corresponding to neutral content.

%% file: results.tex
\section{Results}
\label{sec:results}

\textbf{Classification Results} All combinations of feature extraction models $\mathcal{M}_{d2v}$ and classification functions $h_\theta$ produced a total of $N=289$ possible experts. 
We found that the top 10 highest performing parameter sets across all five balanced training sets were the same for all $\mathcal{M}_{d2v}$. 
{In particular, a maximum distance between the current and predicted word within a sentence of 5, ignoring all words that occurred less than 10 times, and an initial learning rate of 0.025 with a minimum learning rate of 0.00025, resulted in the highest accuracy. Each of these top performing experts were trained for 20 epochs, with a distributed bag-of-words framework. The optimal preprocessing parameters were also the same across the top 10. Each of these used light stop word removal. The optimal document labeling was also the same across these 10 experts, namely the author-group labeling.}
Recall that this labeling scheme tagged each tweet with the authors' unique identifier as well as the known group identification (RG vs RI). 
While the top models had the same training parameters aside from $\lambda$, the models were trained across 5 different training sets, which led  to experts that could differ significantly.

 \begin{table}
   \centering
   \begin{tabular}{cc}
   \begin{tabular}{ccccc}
    \toprule
      \multicolumn{5}{c}{Results for Top Classifier Overall}  \\
      \midrule
         $\gamma$  & Precis.  & Recall & {F1} & Labeled  \\
     \midrule
     0.50 & 0.757 & 0.757 & {0.757} & 100.0\%  \\
     0.65 & 0.837 & 0.837 & {0.837} & 70.29\%  \\
     0.75 & 0.883 & 0.882 & {0.882} & 53.99\%  \\
     0.85 & 0.924 & 0.924 & {0.924} & 37.93\%  \\
     0.95 & 0.970 & 0.970 & {0.970} & 18.49\%   \\
    \bottomrule
   \end{tabular} & 
   \begin{tabular}{ccccc}
      \toprule
         \multicolumn{5}{c}{Results for Panel of Experts} \\
         \midrule
            $\gamma$  & Precis.  & Recall & {F1} & Labeled  \\
        \midrule
        0.50 & 0.763 & 0.762 & {0.762} & 100.0\% \\ 
        0.65 & 0.854 & 0.854 & {0.854} & 66.44\% \\ 
        0.75 & 0.897 & 0.897 & {0.897} & 49.43\% \\ 
        0.85 & 0.939 & 0.939 & {0.939} & 33.45\% \\ 
        0.95 & 0.977 & 0.977 & {0.977} & 15.38\% \\ 
    \bottomrule
      \end{tabular} \\
   \end{tabular}\vspace{0.5\baselineskip}
   
   \caption{Classification scores for the top classifier (left) and a panel of experts using the top 25 experts (right). $\gamma$ is the confidence threshold, and ``Labeled'' is the percentage of examples in the test set that are labeled as either hate or counter at a confidence level of $\gamma$.}
   \label{tab:results}
\end{table}

These top 10 experts had individual F1 scores of $0.755\pm0.0012$ on their individual test sets, when forced to make a classification  (confidence $\gamma=1/2$). Taking into account that each $\mathcal{T}_{out,i}$ was balanced, containing 50,000 hate tweets and 50,000 counter tweets, these F1 scores do not suffer from accuracy inflation that would occur with an unbalanced test set \cite{zhang2019hate,macavaney2019hate}. This result compares well to previous studies that used smaller unbalanced data sets and achieved F1 scores ranging from 0.49 to 0.77 \cite{mathew2019thou, mathew2018analyzing, ziems2020racism}.

As mentioned in Section~\ref{sec:methods}, we did not only use experts in isolation, but also in an ensemble learning approach where the experts could vote on the class label for each tweet in a given test set.  
Due to variations in the training sets and parameters, each expert had a slightly different view of the language, suggesting that combining their knowledge might be beneficial. 
Using the top 10 experts as a panel, instead of individually as just discussed, we obtained an improved average F1 score across all 5 out-of-sample test sets of $0.7616\pm 0.00083$. 
Increasing the size of the panel to include the top 25 experts resulted in an average F1 score across the 5 test sets of $0.7618\pm 0.0007$, see Table~\ref{tab:results}.
We used this large panel for all of our subsequent results.

We also obtained improved results when we varied confidence threshold $\gamma$ and allowed the experts to withhold their vote on contentious tweets.
Not surprisingly, we found that increasing the confidence threshold decreased the number of tweets classified as hate or counter speech.
However, we also found that this led to an increased overall precision, recall, and F1 score, since the labeled tweets were those for which the panel was more certain (see Table~\ref{tab:results}).

 \subsection{Comparison to Human Judgment}
\begin{figure}%
   \centering
   \includegraphics[width=.8\textwidth]{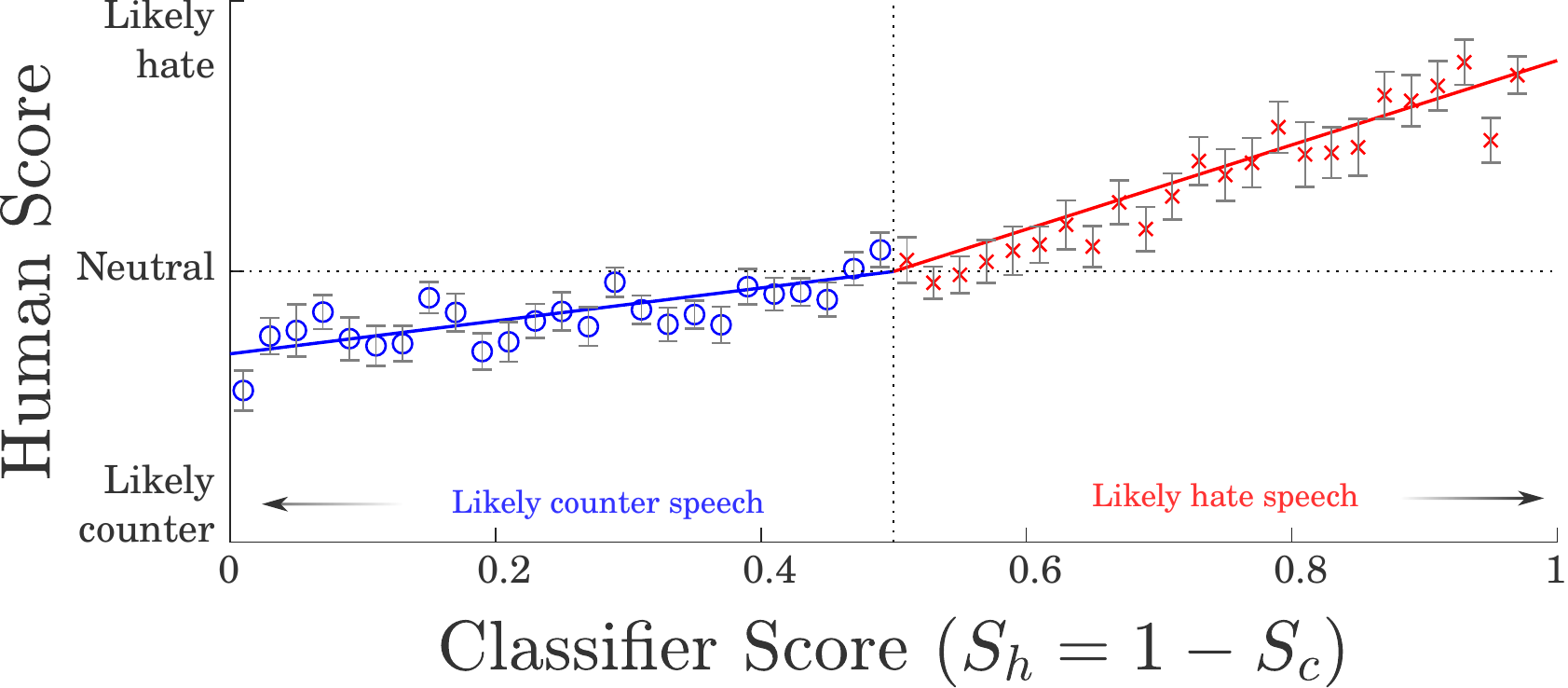}
   \caption{Human judgment of hate and counter speech  corresponds to automated classification (panel of 25 experts).  Average human judgments of tweets classified as counter speech by our method are shown in blue (left-half), and judgments for tweets classified as hate are shown in red (right-half).  Individual human judgments are averaged across bins of width $0.02$ of classifier scores for the original tweet.  Error bars represent $\pm$ one standard error.
   }
   \label{fig:crowdsource}
\end{figure}

 Our crowdsourcing results, shown in Figure~\ref{fig:crowdsource}, suggest that our automated classifier aligns well  with human judgment. Overall correlation between classifier scores and human judgments was $r=0.94$. The correlation was somewhat lower for tweets classified as counter speech ($r=0.75$) than for those classified as hate ($r=0.96$), indicating that to humans counter speech looks more like a `neutral' discourse than hate speech does. As expected, classifier scores around 0.5 received intermediate hate scores from human judges as well.

\subsection{Tree Coloring and Analysis}
 \begin{figure}%
   \centering
   \includegraphics[width=1\textwidth]{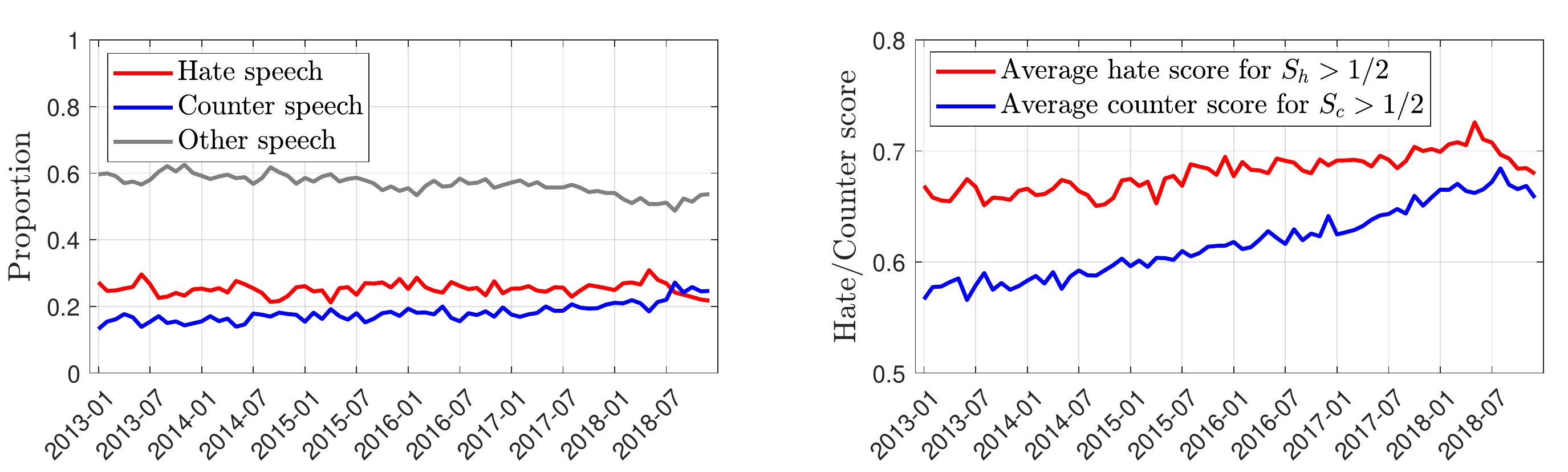}
   \caption{Proportion of hate, counter, and other speech in reply trees from 2013-2018, using a $\gamma=0.75$ threshold (left panel), and average hate and counter score of tweets exceeding the $\gamma=1/2$ threshold (right panel). After the establishment of RI in April 2018, the proportion of counter speech increases, eventually overtaking hate speech around July 2018, and the ongoing increase in polarization is slowed down as indicated by a decrease in average hate and counter scores.}
   \label{fig:frequency}
\end{figure}
  We next used our classifier to label out-of-sample conversations (reply trees) related to current societal and political issues on German Twitter between 2013 and 2018. We asked two questions.
 
 First, how do hate and counter speech develop over time? To study this, we calculated the proportion of hate and counter speech of all speech occurring in each month (using  $\gamma=0.75$), as well as the average hate and counter score for all tweets exceeding $\gamma=1/2$. As Figure~\ref{fig:frequency} shows, the proportion of hate speech was rather stable throughout the examined period, slightly increasing towards the end (red line in the left panel). However, it's extremity was consistently increasing over time (red line in the right panel). The proportion of counter speech was increasing somewhat throughout this period (blue line in the left panel), but its extremity increased quite strongly (blue line in the right panel). A notable change occurred in May 2018, when RI became active: the proportion of counter and other speech increased, and the proportion as well as extremity of hate speech decreased in the following months. This result suggests that organized counter speech can help balance polarized and hateful discourse.

\begin{figure}
\centering
\subfloat[November 2017 - April 2018]{
\includegraphics[width=0.45\textwidth]{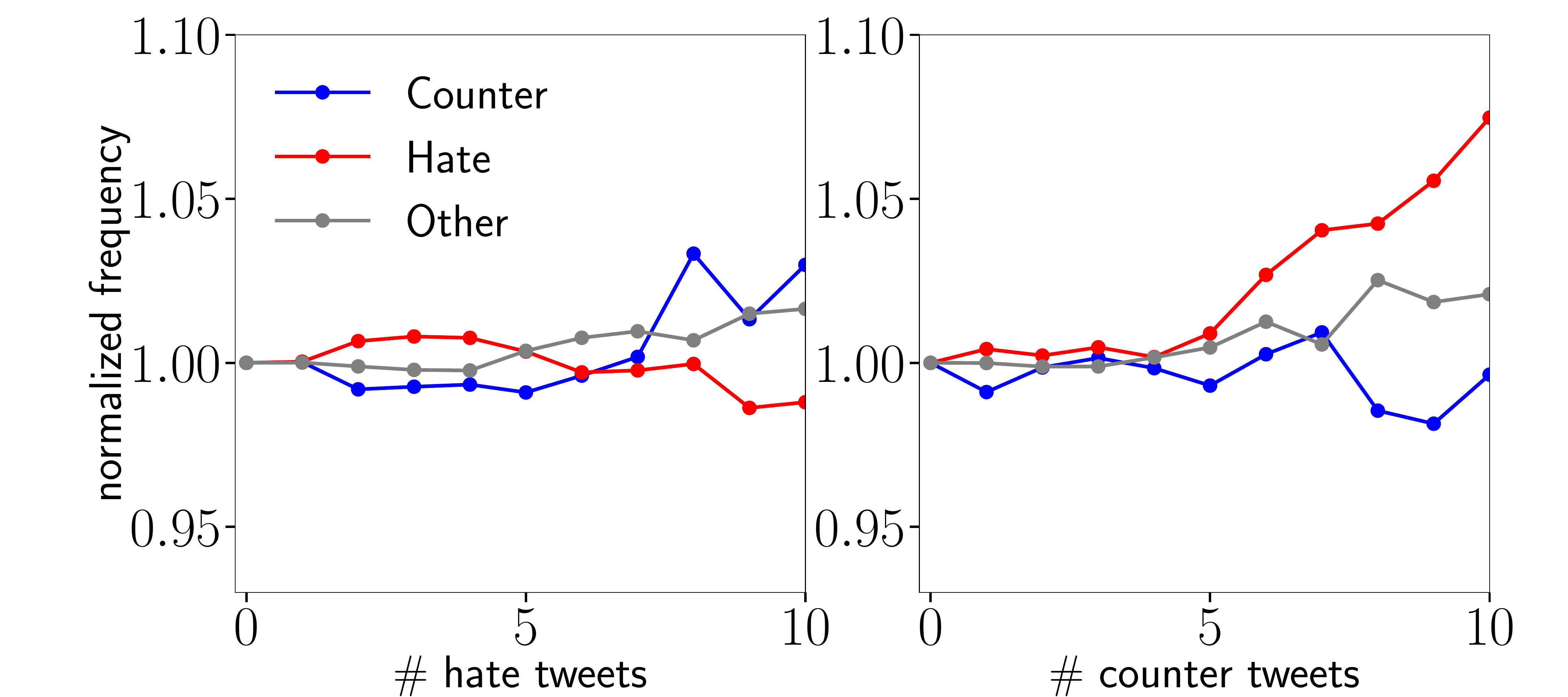}
\label{fig:power1}}
\qquad
\subfloat[May 2018 - October 2018]{
\includegraphics[width=0.45\textwidth]{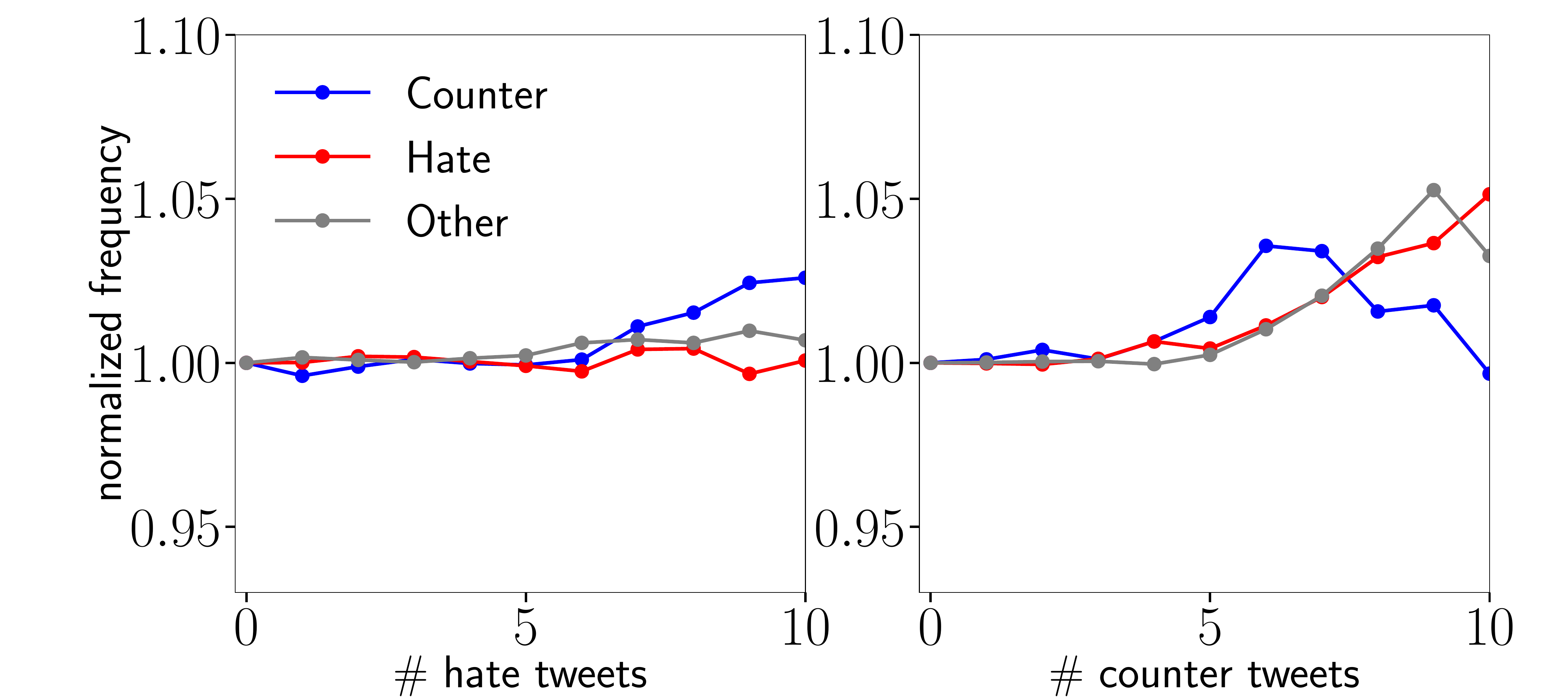}
\label{fig:power2}}
\caption{
Frequency of hate, counter, and other tweets following a hate (counter) tweet, normalized by the overall frequency of these types of tweets in a tree. 
 Panel (a) shows the 6-months period before the establishment of RI, and panel (b) shows the 6-months period after RI was formed. By comparing the right panels of both (a) and (b), tweets from organized counter speech tend to attract more counter and other speech and attract less hate speech than tweets from non-organized counter speech.}
\label{fig:power}
\end{figure}

Second, we conducted an initial analyses into how hate and counter speech interact in reply trees. We asked, how do tweets identified as hate or counter speech change the expected frequency of future hate and counter speech in the reply tree? For this analysis, we used reply trees that have at least 10 tweets identified as hate and at least 10 identified as counter speech, using a 70\% threshold on scores assigned by a panel of the top 25 experts. 
We measured the overall frequency of assigned labels in every individual tree, and tracked how this frequency increases or decreases in time as more tweets identified as hate or counter are posted.
We compared 6-month periods before and after the establishment of RI. Results are shown in Figure~\ref{fig:power}. Before RI was founded, Figure~\ref{fig:power1}, low amount of hate tweets (first panel) somewhat attracted additional hate and suppressed counter speech. However, once many hate tweets were posted, counter speech increased and hate decreased. Similarly, counter tweets (second panel) did not have much effect on hate at first but once there were many counter tweets in a tree they attracted much more hate speech.  
Importantly, counter speech attracted less hate and stimulated additional counter speech more effectively after RI was formed in April 2018 (Figure~\ref{fig:power2}). In all time periods, we also found that counter speech tweets were more likely than hate speech to stimulate neutral or unclassified speech; suggesting that counter speech contributed to depolarizing individual discussions.

Taken together, these results suggest that organized counter speech was associated with a more balanced discourse, reflected in an increased proportion of counter speech in discussions and reduced extremity of hate (Figure~\ref{fig:frequency}) and counter speech having a strong influence in attracting more counter and neutral speech while not attracting more hate (Figure~\ref{fig:power}).

%% file: supp.tex
\newpage\
\beginsupplement
\section*{Supplementary Materials}

\subsection*{Light stop word list}
 als, also, am, an, auf, aus, bei, bis, da, damit,
dann, das, daß, dass, dem, den, der, des, die, dies, ein, eine, ei, nem, einen, einer, eines, einige, einigem, einigen, einiger, einiges, es, im, in, ins, ob, oder, so, sondern, um, und, unter, vom, von, vor, zu, zum, zur'
\subsection*{Supplementary Tables}

\begin{table}[h!]
    \centering
    \begin{tabular}{l:ll}
    \toprule
    & Reconquista Germanica & Reconquista Internet\\
    \midrule
     Members Identified & 2,120& 1,472 \\
     Num. of Tweets & 4,689,294 & 4,323,881 \\
     & &  \\
     &\multirow{3}{*}{{\shortstack[l]{ \attn{\textbf{X}},  QFD,  )))NAME(((,\\  \#shadowbanned,  Böhm Liste,  \\ gab.ai,  \#ichbineinnazi}}} &\multirow{3}{*}{{\shortstack[l]{\#FBPE,  MInT,  (((NAME))), 87\%,  \attn{\textbf{O}}, \\ \#wirsindmehr,  \#noAfD, \#noNazis, \\\#EUFirst, \#FBR, \#OneWorld}}} \\ 
    Bio Features &  \\
    &  \\
    &   \\
    \bottomrule
    \end{tabular}\vspace{0.5\baselineskip}
    \caption{Summary statistics and features of Reconquista Germanica and Reconquista Internet.} \label{tab:summary}
\end{table}

%% file: main.bbl
\begin{thebibliography}{48}
\providecommand{\natexlab}[1]{#1}
\providecommand{\url}[1]{\texttt{#1}}
\expandafter\ifx\csname urlstyle\endcsname\relax
  \providecommand{\doi}[1]{doi: #1}\else
  \providecommand{\doi}{doi: \begingroup \urlstyle{rm}\Url}\fi

\bibitem[Bakalis(2015)]{bakalis2015}
Chara Bakalis.
\newblock \emph{Cyberhate: {A}n issue of continued concern for the council of
  {E}urope's anti-racism commission}.
\newblock Council of Europe, 2015.

\bibitem[Hawdon et~al.(2017)Hawdon, Oksanen, and
  R{\"a}s{\"a}nen]{hawdon2017exposure}
James Hawdon, Atte Oksanen, and Pekka R{\"a}s{\"a}nen.
\newblock Exposure to online hate in four nations: A cross-national
  consideration.
\newblock \emph{Deviant Behav.}, 38\penalty0 (3):\penalty0 254--266, 2017.

\bibitem[Oksanen et~al.(2018)Oksanen, Kaakinen, Minkkinen, R{\"a}s{\"a}nen,
  Enjolras, and Steen-Johnsen]{oksanen2018perceived}
Atte Oksanen, Markus Kaakinen, Jaana Minkkinen, Pekka R{\"a}s{\"a}nen, Bernard
  Enjolras, and Kari Steen-Johnsen.
\newblock Perceived societal fear and cyberhate after the {November} 2015 paris
  terrorist attacks.
\newblock \emph{Terror. Political Violence}, pages 1--20, 2018.

\bibitem[M{\"u}ller and Schwarz(2019)]{muller2019fanning}
Karsten M{\"u}ller and Carlo Schwarz.
\newblock Fanning the flames of hate: {Social} media and hate crime.
\newblock \emph{SSRN:3082972}, 2019.

\bibitem[{\'A}lvarez-Benjumea and Winter(2018)]{alvarez2018normative}
Amalia {\'A}lvarez-Benjumea and Fabian Winter.
\newblock Normative change and culture of hate: An experiment in online
  environments.
\newblock \emph{Eur. Sociol. Rev.}, 34\penalty0 (3):\penalty0 223--237, 2018.

\bibitem[Chandrasekharan et~al.(2017)Chandrasekharan, Pavalanathan, Srinivasan,
  Glynn, Eisenstein, and Gilbert]{chandrasekharan2017you}
Eshwar Chandrasekharan, Umashanthi Pavalanathan, Anirudh Srinivasan, Adam
  Glynn, Jacob Eisenstein, and Eric Gilbert.
\newblock You can't stay here: The efficacy of {Reddit}'s 2015 ban examined
  through hate speech.
\newblock In \emph{Proceedings of the {ACM} on Human-Computer Interaction},
  volume~1, pages 1--22, 2017.

\bibitem[Benesch et~al.(2016)Benesch, Ruths, Dillon, Saleem, and
  Wright]{benesch2016considerations}
S~Benesch, D~Ruths, KP~Dillon, H~M Saleem, and L~Wright.
\newblock Considerations for successful counterspeech, 2016.
\newblock URL
  \url{https://https://dangerousspeech.org/considerations-for-successful-counterspeech/}.

\bibitem[Rieger et~al.(2018)Rieger, Schmitt, and Frischlich]{rieger2018hate}
Diana Rieger, Josephine~B Schmitt, and Lena Frischlich.
\newblock Hate and counter-voices in the internet: Introduction to the special
  issue.
\newblock \emph{{SCM} Stud. Commun. Media}, 7\penalty0 (4):\penalty0 459--472,
  2018.

\bibitem[Gaffney et~al.(2019)Gaffney, Farrington, Espelage, and
  Ttofi]{gaffney2019cyberbullying}
Hannah Gaffney, David~P Farrington, Dorothy~L Espelage, and Maria~M Ttofi.
\newblock Are cyberbullying intervention and prevention programs effective? {A}
  systematic and meta-analytical review.
\newblock \emph{Aggress. Violent Behav.}, 45:\penalty0 134--153, 2019.

\bibitem[Gagliardone et~al.(2015)Gagliardone, Gal, Alves, and
  Martinez]{gagliardone2015countering}
Iginio Gagliardone, Danit Gal, Thiago Alves, and Gabriela Martinez.
\newblock \emph{Countering online hate speech}.
\newblock Unesco Publishing, 2015.

\bibitem[Mathew et~al.(2018)Mathew, Kumar, Goyal, and
  Mukherjee]{mathew2018analyzing}
Binny Mathew, Navish Kumar, Pawan Goyal, and Animesh Mukherjee.
\newblock Analyzing the hate and counter speech accounts on {Twitter}.
\newblock \emph{arXiv:1812.02712}, 2018.

\bibitem[Mathew et~al.(2019)Mathew, Saha, Tharad, Rajgaria, Singhania, Maity,
  Goyal, and Mukherjee]{mathew2019thou}
Binny Mathew, Punyajoy Saha, Hardik Tharad, Subham Rajgaria, Prajwal Singhania,
  Suman~Kalyan Maity, Pawan Goyal, and Animesh Mukherjee.
\newblock Thou shalt not hate: {Countering} online hate speech.
\newblock In \emph{Proceedings of the International {AAAI} Conference on Web
  and Social Media}, volume~13, pages 369--380, 2019.

\bibitem[Wright et~al.(2017)Wright, Ruths, Dillon, Saleem, and
  Benesch]{wright2017vectors}
Lucas Wright, Derek Ruths, Kelly~P Dillon, Haji~Mohammad Saleem, and Susan
  Benesch.
\newblock Vectors for counterspeech on {Twitter}.
\newblock In \emph{Proceedings of the first workshop on abusive language
  online}, pages 57--62, 2017.

\bibitem[Ziegele et~al.(2018)Ziegele, Jost, Bormann, and
  Heinbach]{ziegele2018journalistic}
Marc Ziegele, Pablo Jost, Marike Bormann, and Dominique Heinbach.
\newblock Journalistic counter-voices in comment sections: {Patterns},
  determinants, and potential consequences of interactive moderation of uncivil
  user comments.
\newblock \emph{{SCM} Stud. Commun. Media}, 7\penalty0 (4):\penalty0 525--554,
  2018.

\bibitem[Ziems et~al.(2020)Ziems, He, Soni, and Kumar]{ziems2020racism}
Caleb Ziems, Bing He, Sandeep Soni, and Srijan Kumar.
\newblock Racism is a virus: Anti-asian hate and counterhate in social media
  during the {COVID}-19 crisis.
\newblock \emph{arXiv:2005.12423}, 2020.

\bibitem[Blaya(2019)]{blaya2019cyberhate}
Catherine Blaya.
\newblock Cyberhate: A review and content analysis of intervention strategies.
\newblock \emph{Aggress. Violent Behav.}, 45:\penalty0 163--172, 2019.

\bibitem[Weber(2009)]{weber2009manual}
Anne Weber.
\newblock \emph{Manual on hate speech}.
\newblock Council Of Europe, 2009.

\bibitem[you(2019)]{youtube_guide}
Youtube: {Hate} speech policy, 2019.
\newblock URL \url{https://support.google.com/youtube/answer/2801939}.

\bibitem[twi(2019)]{twitter_guide}
{Twitter}: {Hateful} conduct policy, 2019.
\newblock URL
  \url{https://help.twitter.com/en/rules-and-policies/hateful-conduct-policy}.

\bibitem[fac(2019)]{facebook_guide}
{Facebook}: {Hate} speech, 2019.
\newblock URL \url{https://www.facebook.com/communitystandards/hate_speech}.

\bibitem[ser(2019)]{seriously-website}
Seriously, 2019.
\newblock URL \url{http://www.www.seriously.org}.

\bibitem[smh(2019)]{smh-website}
Social media helpline, 2019.
\newblock URL
  \url{https://socialmediahelpline.com/counterspeech-dos-and-donts-for-students/}.

\bibitem[Habermas(2015)]{habermas2015between}
J{\"u}rgen Habermas.
\newblock \emph{Between facts and norms: {C}ontributions to a discourse theory
  of law and democracy}.
\newblock John Wiley \& Sons, 2015.

\bibitem[Brassard-Gourdeau and Khoury(2018)]{brassard2018impact}
{\'E}loi Brassard-Gourdeau and Richard Khoury.
\newblock Impact of sentiment detection to recognize toxic and subversive
  online comments.
\newblock \emph{arXiv:1812.01704}, 2018.

\bibitem[Burnap et~al.(2015)Burnap, Rana, Avis, Williams, Housley, Edwards,
  Morgan, and Sloan]{burnap2015detecting}
Pete Burnap, Omer~F Rana, Nick Avis, Matthew Williams, William Housley, Adam
  Edwards, Jeffrey Morgan, and Luke Sloan.
\newblock Detecting tension in online communities with computational {Twitter}
  analysis.
\newblock \emph{Technol. Forecast. Soc. Change}, 95:\penalty0 96--108, 2015.

\bibitem[Burnap and Williams(2016)]{burnap2016us}
Pete Burnap and Matthew~L Williams.
\newblock Us and them: identifying cyber hate on {Twitter} across multiple
  protected characteristics.
\newblock \emph{{EPJ} Data science}, 5\penalty0 (1):\penalty0 11, 2016.

\bibitem[Ribeiro et~al.(2018)Ribeiro, Calais, Santos, Almeida, and
  Meira~Jr]{ribeiro2018characterizing}
Manoel~Horta Ribeiro, Pedro~H Calais, Yuri~A Santos, Virg{\'\i}lio~AF Almeida,
  and Wagner Meira~Jr.
\newblock Characterizing and detecting hateful users on {Twitter}.
\newblock In \emph{Twelfth international {AAAI} conference on web and social
  media}, 2018.

\bibitem[Zhang and Luo(2019)]{zhang2019hate}
Ziqi Zhang and Lei Luo.
\newblock Hate speech detection: A solved problem? the challenging case of long
  tail on {Twitter}.
\newblock \emph{Semantic Web}, 10\penalty0 (5):\penalty0 925--945, 2019.

\bibitem[Bosco et~al.(2018)Bosco, Felice, Poletto, Sanguinetti, and
  Maurizio]{bosco2018overview}
Cristina Bosco, Dell'Orletta Felice, Fabio Poletto, Manuela Sanguinetti, and
  Tesconi Maurizio.
\newblock Overview of the evalita {EVALITA} hate speech detection task.
\newblock In \emph{{EVALITA} 2018-Sixth Evaluation Campaign of Natural Language
  Processing and Speech Tools for {Italian}}, volume 2263. CEUR, 2018.

\bibitem[de~Gibert et~al.(2018)de~Gibert, Perez, Garc{\'\i}a-Pablos, and
  Cuadros]{gibert2018hate}
Ona de~Gibert, Naiara Perez, Aitor Garc{\'\i}a-Pablos, and Montse Cuadros.
\newblock Hate speech dataset from a white supremacy forum.
\newblock In \emph{Proceedings of the 2nd Workshop on Abusive Language Online
  ({ALW}2)}, pages 11--20. Association for Computational Linguistics, 2018.

\bibitem[Kshirsagar et~al.(2018)Kshirsagar, Cukuvac, McKeown, and
  McGregor]{kshirsagar2018predictive}
Rohan Kshirsagar, Tyus Cukuvac, Kathleen McKeown, and Susan McGregor.
\newblock Predictive embeddings for hate speech detection on {Twitter}.
\newblock In \emph{Proceedings of the 2nd Workshop on Abusive Language Online
  ({ALW}2)}, pages 26--32. Association for Computational Linguistics, 2018.

\bibitem[MacAvaney et~al.(2019)MacAvaney, Yao, Yang, Russell, Goharian, and
  Frieder]{macavaney2019hate}
Sean MacAvaney, Hao-Ren Yao, Eugene Yang, Katina Russell, Nazli Goharian, and
  Ophir Frieder.
\newblock Hate speech detection: Challenges and solutions.
\newblock \emph{PLOS ONE}, 14\penalty0 (8), 2019.

\bibitem[Malmasi and Zampieri(2018)]{malmasi2018challenges}
Shervin Malmasi and Marcos Zampieri.
\newblock Challenges in discriminating profanity from hate speech.
\newblock \emph{J. Exp. Theor. Artif. Intell.}, 30\penalty0 (2):\penalty0
  187--202, 2018.

\bibitem[Pitsilis et~al.(2018)Pitsilis, Ramampiaro, and
  Langseth]{pitsilis2018effective}
Georgios~K Pitsilis, Heri Ramampiaro, and Helge Langseth.
\newblock Effective hate-speech detection in {Twitter} data using recurrent
  neural networks.
\newblock \emph{Appl. Intell.}, 48\penalty0 (12):\penalty0 4730--4742, 2018.

\bibitem[Al-Hassan and Al-Dossari(2019)]{al2019detection}
Areej Al-Hassan and Hmood Al-Dossari.
\newblock Detection of hate speech in social networks: a survey on multilingual
  corpus.
\newblock In \emph{6th International Conference on Computer Science and
  Information Technology}, 2019.

\bibitem[Vidgen and Yasseri(2020)]{vidgen2020detecting}
Bertie Vidgen and Taha Yasseri.
\newblock Detecting weak and strong islamophobic hate speech on social media.
\newblock \emph{J. Inf. Technol. Politics}, 17\penalty0 (1):\penalty0 66--78,
  2020.

\bibitem[Zimmerman et~al.(2018)Zimmerman, Kruschwitz, and
  Fox]{zimmerman2018improving}
Steven Zimmerman, Udo Kruschwitz, and Chris Fox.
\newblock Improving hate speech detection with deep learning ensembles.
\newblock In \emph{Proceedings of the Eleventh International Conference on
  Language Resources and Evaluation (LREC 2018)}, 2018.

\bibitem[Le and Mikolov(2014)]{doc2vec}
Quoc Le and Tomas Mikolov.
\newblock Distributed representations of sentences and documents.
\newblock In \emph{International conference on machine learning}, pages
  1188--1196, 2014.

\bibitem[Pennington et~al.(2014)Pennington, Socher, and
  Manning]{pennington2014glove}
Jeffrey Pennington, Richard Socher, and Christopher~D Manning.
\newblock Glove: {Global} vectors for word representation.
\newblock In \emph{Proceedings of the 2014 conference on empirical methods in
  natural language processing}, pages 1532--1543, 2014.

\bibitem[Devlin et~al.(2019)Devlin, Chang, Lee, and Toutanova]{devlin2018bert}
Jacob Devlin, Ming-Wei Chang, Kenton Lee, and Kristina Toutanova.
\newblock {BERT}: Pre-training of deep bidirectional transformers for language
  understanding.
\newblock In \emph{Proceedings of NAACL-HLT 2019}, pages 4171–--4186, 2019.

\bibitem[Davidson et~al.(2017)Davidson, Warmsley, Macy, and
  Weber]{davidson2017automated}
Thomas Davidson, Dana Warmsley, Michael Macy, and Ingmar Weber.
\newblock Automated hate speech detection and the problem of offensive
  language.
\newblock In \emph{Eleventh international {AAAI} conference on {Web} and social
  media}, 2017.

\bibitem[Kennedy et~al.(2017)Kennedy, McCollough, Dixon, Bastidas, Ryan, Loo,
  and Sahay]{kennedy2017technology}
George Kennedy, Andrew McCollough, Edward Dixon, Alexei Bastidas, John Ryan,
  Chris Loo, and Saurav Sahay.
\newblock Technology solutions to combat online harassment.
\newblock In \emph{Proceedings of the first workshop on abusive language
  online}, pages 73--77, 2017.

\bibitem[Schmidt and Wiegand(2017)]{schmidt2017survey}
Anna Schmidt and Michael Wiegand.
\newblock A survey on hate speech detection using natural language processing.
\newblock In \emph{Proceedings of the Fifth International Workshop on Natural
  Language Processing for Social Media}, pages 1--10, 2017.

\bibitem[{\v R}eh{\r u}{\v r}ek and Sojka(2010)]{rehurek_lrec}
Radim {\v R}eh{\r u}{\v r}ek and Petr Sojka.
\newblock Software framework for topic modelling with large corpora.
\newblock In \emph{Proceedings of the {LREC} 2010 Workshop on New Challenges
  for {NLP} Frameworks}, pages 45--50, Valletta, Malta, 2010. ELRA.

\bibitem[Lau and Baldwin(2016)]{lau2016empirical}
Jey~Han Lau and Timothy Baldwin.
\newblock An empirical evaluation of doc2vec with practical insights into
  document embedding generation.
\newblock In \emph{Proceedings of the 1st Workshop on Representation Learning
  for {NLP}}, pages 78--86. Association for Computational Linguistics, 2016.

\bibitem[Pedregosa et~al.(2011)Pedregosa, Varoquaux, Gramfort, Michel, Thirion,
  Grisel, Blondel, Prettenhofer, Weiss, Dubourg, Vanderplas, Passos,
  Cournapeau, Brucher, Perrot, and Duchesnay]{scikit-learn}
F.~Pedregosa, G.~Varoquaux, A.~Gramfort, V.~Michel, B.~Thirion, O.~Grisel,
  M.~Blondel, P.~Prettenhofer, R.~Weiss, V.~Dubourg, J.~Vanderplas, A.~Passos,
  D.~Cournapeau, M.~Brucher, M.~Perrot, and E.~Duchesnay.
\newblock Scikit-learn: Machine learning in {P}ython.
\newblock \emph{J. Mach. Learn. Res.}, 12:\penalty0 2825--2830, 2011.

\bibitem[Jaki and De~Smedt(2019)]{jaki2019right}
Sylvia Jaki and Tom De~Smedt.
\newblock Right-wing {German} hate speech on {Twitter}: Analysis and automatic
  detection.
\newblock \emph{arXiv:1910.07518}, 2019.

\bibitem[Chen and Guestrin(2016)]{chen2016xgboost}
Tianqi Chen and Carlos Guestrin.
\newblock {xgboost}: A scalable tree boosting system.
\newblock In \emph{Proceedings of the 22nd {ACM} {SIGKDD} international
  conference on knowledge discovery and data mining}, pages 785--794, 2016.

\end{thebibliography}
